\documentclass[reprint,amsmath,amssymb,aps,superscriptaddress,twocolumn,float]{revtex4-1}
\usepackage{graphicx}
\usepackage{graphics}
\usepackage{dcolumn}
\usepackage{bm}
\usepackage{amsfonts}
\usepackage{amssymb}
\usepackage{float}
\usepackage{xcolor}
\begin{document}
\title{Anisotropic spin Hall and spin Nernst effects in bismuth semimetal}

\author{Guang-Yu Guo}
\email{gyguo@phys.ntu.edu.tw}
\affiliation{Department of Physics, National Taiwan University, Taipei 10617, Taiwan\looseness=-1}
\affiliation{Physics Division, National Center for Theoretical Sciences, Taipei 10617, Taiwan\looseness=-1}

\date{\today}

\begin{abstract}

Bismuth is an archetypal semimetal with gigantic spin-orbit coupling 
and it has been a major source material for the discovery of seminal phenomena in solid state physics 
for more than a century. In recent years, spin current transports in bismuth
have also attracted considerable attention. In this paper, we theoretically study 
both spin Hall effect (SHE) and spin Nernst effect (SNE) in bismuth,
based on relativistic band structure calculations.
First, we find that there are three independent tensor elements of
spin Hall conductivity (SHC) ($\sigma_{ij}^s$) and spin Nernst conductivity (SNC)
($\alpha_{ij}^s$), namely, $Z_{yx}^z$, $Z_{xz}^y$, and $Z_{zy}^x$, where $Z = \sigma$ or $\alpha$.
We calculate all the elements as a function of the Fermi energy.
Second, we find that all SHC elements are large, being $\sim$1000 ($\hbar$/e)(S/cm).
Furthermore, because of its low electrical
conductivity, the spin Hall angles are gigantic, being $\sim$20 \%.
Third, all the calculated SNC elements are also pronounced, being
comparable to that [$\sim$0.13 ($\hbar$/e)(A/m-K)] of gold.
Finally, in contrast to Pt and Au where $Z_{yx}^z = Z_{xz}^y = Z_{zy}^x$, 
the SHE and SNE in bismuth are anisotropic, i.e., $Z_{yx}^z \ne Z_{xz}^y \ne Z_{zy}^x$. 
In particular, SNC is highly anisotropic, and $\alpha_{yx}^z$, $\alpha_{xz}^y$ and $\alpha_{zy}^x$ 
differ even in sign. Also, such anisotropy in SHE can be significantly enhanced
by either electron or hole doping.
Consequently, the Hall voltages due to the inverse SHE and inverse SNE from the different 
conductivity elements could cancel each other and thus result
in a small spin Hall angle if polycrystalline samples are used, which
may explain why the measured spin Hall angles ranging from nearly 0 to 25 \% have been reported. 
We hope that these interesting findings would stimulate further experiments on bismuth
using highly oriented single crystal specimens.

\end{abstract}

\maketitle

\section{INTRODUCTION}

In the past two decades, spin transport electronics (spintronics) has attracted considerable attention
because of its promising applications in data storage and processing and other electronic
technologies. Spin current generation, detection and manipulation are key issues in spintronics.
Spin Hall effect (SHE), first proposed in 1971~\cite{Dyakonov1971}, refers to the transverse 
spin current generation in a nonmagnetic solid with 
relativistic spin-orbit coupling (SOC) by an electric 
field~\cite{Dyakonov1971,Hirsch1999,Murakami2003,Sinova2004,Kato2004,Guo2005,Saitoh2006,Valenzuela2006,Kimura2007,Guo2008,Tanaka2008,Liu2012,Hoffmann2013,Sinova2015}.
It offers a rather unique method for pure spin current generation and manipulation 
without the need of an applied magnetic field 
or a magnetic material. Large SHE has been predicted and also observed in 5$d$ transition metals such as 
platinum and tungsten with strong SOC~\cite{Kimura2007,Guo2008,Tanaka2008,Liu2012,Hoffmann2013,Sinova2015}
In the inverse SHE~\cite{Saitoh2006}, pure spin current is converted
into a transverse charge current in a material with large SOC.
ISHE has been widely used to detect spin currents~\cite{Saitoh2006,Valenzuela2006,Hoffmann2013,Sinova2015}
Similarly, pure transverse spin current could also be generated in a solid by applying
a temperature gradient ($\nabla T$) rather than an electric field~\cite{Cheng2008}. 
This thermally driven spin current generation is called spin Nernst effect (SNE) \cite{Cheng2008}
and would make spintronics powered by heat possible, thus opening a new field known as
spin caloritronics~\cite{Bauer2012}. 
Large SNE has been  recently observed in such materials as Pt film \cite{Meyer2017} 
and tungsten metal \cite{Sheng2017}. 

Bismuth (Bi) crystallizes in a rhombohedral $R\bar{3}$m structure at ambient conditions~\cite{Cucka1962}
and is an archetypal semimetal~\cite{Gallo1963,Liu1995}. 
Bismuth is the heaviest nonradioative element in the periodic table, and thus Bi 
has a gigantic SOC effect~\cite{Liu1995} (see also Fig. 1 below). 
For more than a century, Bi has been a key source material for the discovery of important phenomena 
in solid state physics, such as the large thermoelectric effect~\cite{Seebeck1825,Gallo1963},
the largest Hall coefficient and highest magnetoresistance~\cite{Kamerlingh-Onnes1912,Kapitza1928}.
The latter two unusual properties have been attributed to the large SOC in Bi~\cite{Fuseya2015}.
More recently, Bi has also been found to host intrinsic superconductivity~\cite{Prakash2017} and
to be a higher order topological insulator~\cite{Schindler2018} and also a rotational symmetry protected
topological crystalline insulator~\cite{Hsu2019}.
Because of its strong SOC, one would expect bismuth to exhibit large SHE and SNE.
Indeed, the contribution of the electron carriers to the spin Hall conductivity (SHC) 
in bismuth has been studied by model Dirac Hamiltonian and also anisotropic Wolff 
Hamiltonian~\cite{Fuseya2012,Fuseya2014,Fuseya2015b,Chi2022} and
the SHC was estimated to be two-orders of magnitude larger than that of Pt at room temperature.
Furthermore, large SHC of $\sim$470 ($\hbar$/e)(S/cm)
in Bi was predicted by a realistic tight-binding Hamiltonian calculation~\cite{Sahin2015}, 
which is about one quarter of that of platinum ~\cite{Guo2008}.
Since bismuth is a semimetal with the electrical conductivity being much lower than that of platinum~\cite{Gallo1963}, 
one would expect bismuth to have a very large spin Hall angle ($\Theta_{sH}$) (the ratio of the spin current to
the longitudinal charge current, i.e., charge-spin conversion efficiency). 
Therefore, many investigations on spin current phenomena in 
Bi semimetal systems have been carried~\cite{Hou2012,Emoto2016,Zhang2015,Yue2018,Yue2021,Sangiao2021}
and indeed large spin Hall angles of about 10 \% have been reported in some of these experiments~\cite{Hou2012}.

However, the magnitude of the reported $\Theta_{sH}$ values varies from zero (negligibly small) 
to 24 \%, depending on the spin current detection method and also the magnetic material used to inject 
spin current in the experiments~\cite{Yue2021}. Moreover, the signs of the reported $\Theta_{sH}$ values
could also differ. This unsatisfying state of the matter 
is partly due to the fact that apart from the SHE, spin current can also be generated by
other means such as spin pumping at ferromagnetic resonance, SNE, lateral spin valve 
and longitudinal spin Seebeck effect (see \cite{Yue2018} and references therein).
On the other hand, we notice that the SHC (and also spin Nernst conductivity) for a solid is a third-order tensor 
and thus has 27 matrix elements. Of course, many of the elements are zero due to 
the crystalline symmetry constraints. For example, bcc W and fcc Pt has only one independent
nonzero element ($\sigma_{yx}^z$). Nevertheless, it was reported that for hcp metals, 
there are two independent elements ($\sigma_{yx}^z$ and $\sigma_{xz}^y$)~\cite{Freimuth2010}.
These hcp metals could exhibit large anisotropic SHE if $\sigma_{yx}^z$ and $\sigma_{xz}^y$
differ considerably, e.g., in hcp Sc, hcp Zn and hcp Zr~\cite{Freimuth2010}.
Consequently, the Hall voltages generated by the inverse SHE due to different SHC
tensor elements could partially cancel each other and hence may result in an averaged smaller $\Theta_{sH}$ value
if polycrystalline samples are used in the experiments.
As will be discussed below in Sec. II B, rhombohedral bismuth has 4 independent nonzero elements 
($\sigma_{yx}^z$, $\sigma_{xz}^y$, $\sigma_{zy}^x$ and $\sigma_{xx}^y$) (see Table I below). However,
only $\sigma_{yx}^z$ of bismuth was reported in ~\cite{Sahin2015}. Clearly, the knowledge of 
all the four independent SHC tensor elements would help understand the wide spectrum of the reported $\Theta_{sH}$ values
and also guide the development of spintronic devices using Bi semimetal.

Because of its semimetallic nature, the thermoelectric effect in bismuth has also been intensively
investigated and large Seebeck coefficients of about 100 $\mu$V/K were reported~\cite{Seebeck1825,Gallo1963}.
One thus would wonder bismuth may also exhibit large SNE since it has large SOC.  
However, no theoretical nor experimental study on the SNE in bismuth has been reported so far.
Therefore, we have carried out a systematic first-principles density functional theory (DFT) study 
on all the independent nonzero elements of the SHC and also the spin Nernst conductivity (SNC) of bismuth.  
In this paper, we present the main results of this systematic theoretical investigation.
The rest of this paper is organized as follows.
In the next section, we briefly describe the Berry phase formalism for
calculating the intrinsic SHC and SNC of bismuth as well as the computational details.
Section III consists of three subsections.
We first present the calculated relativistic band structure of bismuth
and also analyze its main features in Sec. III A.
We then report the calculated spin Hall conductivity and corresponding estimated spin Hall angle in Sec. III B.
We finally present the calculated spin Nernst conductivity and its temperature dependence as well as
the estimated spin Nernst angles in Sec. III C.
Finally, the conclusions drawn from this work are summarized in Sec. IV.

\section{THEORY AND COMPUTATIONAL DETAILS}

Bismuth crystallizes in a rhombohedral $R\bar{3}$m structure (space group 166) with
two Bi atoms per unit cell.~\cite{Cucka1962} 
The experimental lattice constants and atomic positions measured at room 
temperature~\cite{Cucka1962} are used in the present calculations. 
The electronic band structures are calculated based on the DFT with 
the generalized gradient approximation (GGA) to the exchange-correlation potential \cite{Perdew1996}.
The accurate full-potential projecter-augmented wave  method\cite{Blochl1994},
as implemented in the  Vienna \textit{ab initio} simulation package (VASP) \cite{Kresse1996,Kresse1993},
is used. A large plane-wave cutoff energy of 400 eV is used in all the calculations.   
The self-consistent relativistic band structure calculation is performed with a $\Gamma$-centered $k$-point mesh 
of 16$\times$16$\times$16 used in the  Brillouin zone (BZ) integration by the tetrahedron method \cite{Temmerman1989}.
All the calculations are carried out with an energy convergence within 10$^{-7}$ eV between the 
successive iteration.

The intrinsic SHC is calculated within the Berry phase formalism \cite{Guo2008,Xiao2010}. 
In this approach, the SHC ($\sigma_{ij}^s=J_i^s/E_j$) is simply given by the BZ integration of the spin Berry 
curvature for all occupied bands~\cite{Guo2008},
\begin{equation}
\label{eq:1}
\begin{aligned}
\sigma_{ij}^{s}=e\sum_{n}\int_{BZ}\frac{d\textbf{k}}{(2\pi)^3}f_{\textbf{k}n}\Omega_{ij}^{n,s}(\textbf{k})
\end{aligned}
\end{equation}
\begin{equation}
\label{eq:2} 
\begin{aligned}
\Omega_{ij}^{n,s}=\sum_{n'\neq n}\frac{2Im\left[\langle\textbf{k}n|\{\tau_{s},v_{i}\}/4|\textbf{k}n'\rangle\langle\textbf{k}n'|v_{j}|\textbf{k}n\rangle \right] }{(\epsilon_{\textbf{k}n}-\epsilon_{\textbf{k}n'})^2}
\end{aligned}
\end{equation}
where $J_i^s$ is the $i$th component of spin current density, $f_{\textbf{k}n}$ is the Fermi function, 
and $\Omega_{i,j}^{n,s} (\textbf{k})$ 
is the spin Berry curvature for the $n^{th}$ band at $\textbf{k}$. $i,j=x,y,z$ and $i\neq j$. 
$s$ denotes the spin direction, $\tau_s$ is the Pauli matrix, and $v_i$ is the velocity operator~\cite{Guo2005}.
Similarly, the spin Nernst conductivity ($\alpha_{ij}^s=-J_i^s/\nabla_jT$) can be written as~\cite{Xiao2010,Guo2017}
\begin{equation}
\label{eq:3} 
\begin{aligned}
\alpha_{ij}^{s}&= -\frac{1}{T}\sum_{n}\int_{BZ}\frac{d\textbf{k}}{(2\pi)^3}\Omega_{i,j}^{n,s} (\textbf{k})\\
&\times [(\epsilon_{\textbf{k}n}-\mu)f_{\textbf{k}n}+k_{B}T\ln (1+e^{-\beta(\epsilon_{\textbf{k}n}-\mu)})].
\end{aligned}
\end{equation}
In the SHC and SNC calculations, the velocity $\langle\textbf{k}n'|v_{i}|\textbf{k}n\rangle$ 
and spin velocity $\langle\textbf{k}n|\{\tau_{s},v_{i}\}/4]|\textbf{k}n'\rangle$ matrix elements 
are first calculated from the self-consistent relativistic band structures 
within the projector-augmented wave formalism~\cite{Adolph01} by using the VASP program.
A homemade program~\cite{Guo2014,Guo2017} is then used to calculate the spin Berry curvature [Eq. (2)] 
and also to perform the band summation and the BZ integration with the tetrahedron method [Eqs. (1) and (3)].
A very fine $k$-point mesh of about 124000 $k$-points in the BZ wedge is used, and  
this corresponds to the division of the $\Gamma K$ line into $n_d=40$ intervals. 
Further test calculatons using denser $k$-point meshes of $n_d=50$ and 60 show 
that the calculated SHC and SNC converge within a few percent.

\section{RESULTS AND DISCUSSION}

\begin{figure}[tbph] \centering
\includegraphics[width=8.0cm]{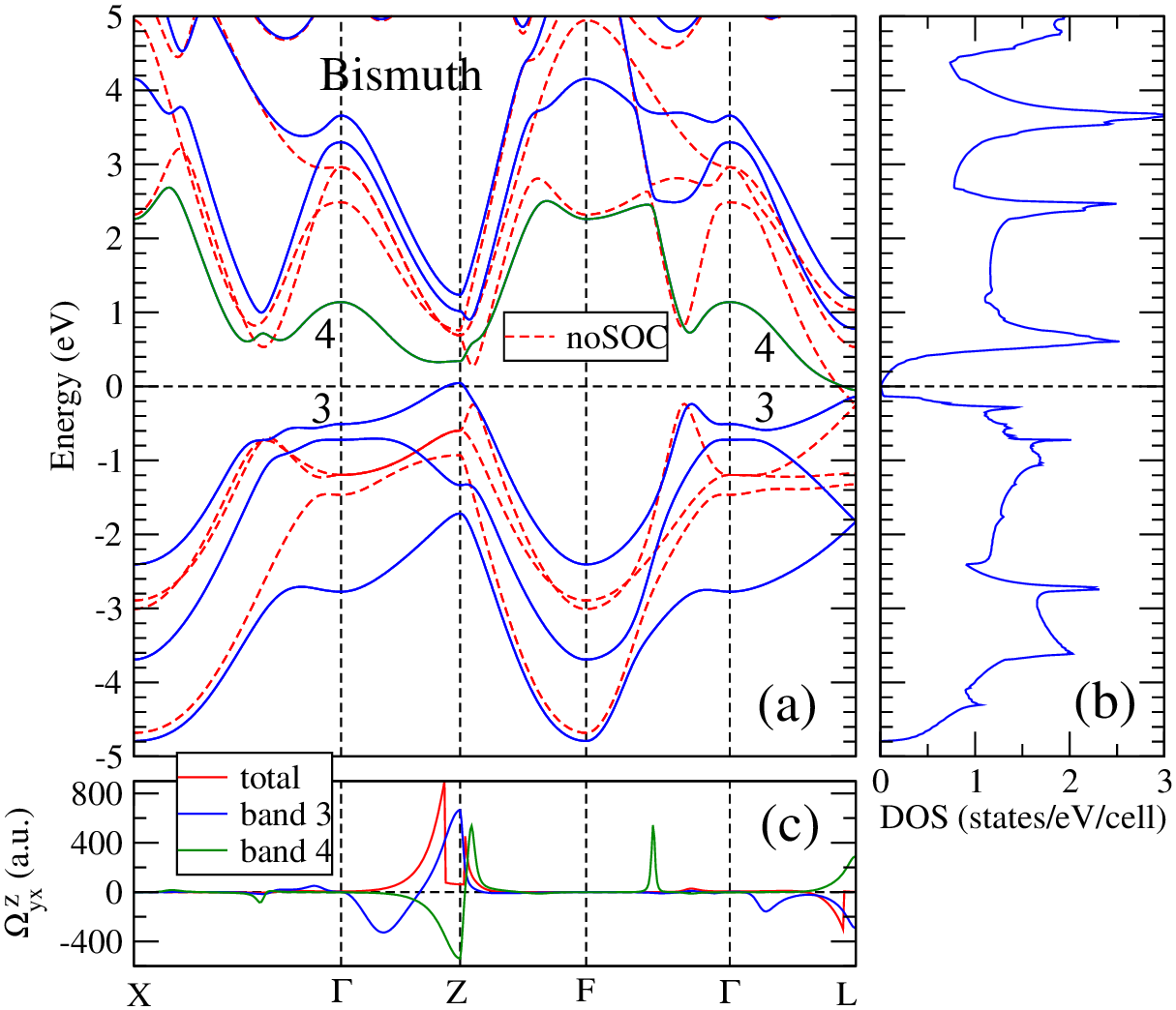}
\caption{(a) Relativisitc band strucuture, (b) total density of states (DOS)
and (c) total and band-resolved spin Berry curvatures ($\Omega_{yx}^z$) 
of bismuth semimetal. In (a), the scalar-relativistic band structure [i.e., without
the spin-orbit coupling (SOC)] is also displayed for comparison (dashed red lines).
The Fermi level is at 0 eV.
}
\label{fig:BS}
\end{figure}

\subsection{Electronic band structures}
The calculated relativistic band structure and total density of states (DOSs) of bismuth are shown in 
Fig. \ref{fig:BS}(a) and Fig. \ref{fig:BS}(b), respectively. 
As reported earlier~\cite{Liu1995,Sahin2015}, bismuth is a semimetal with a small DOS of 0.012 states/eV/cell 
at the Fermi level ($E_F$). Its Fermi surface consists of a small hole pocket at the Z point and
three small electron pockets at the L point [Fig. \ref{fig:BS}(a)]. 
Top valence bands are made up of three broad Kramers-degenerate Bi $p$-orbital dominated bonding bands
while bottom conduction bands consists of three broad Kramers-degenerate Bi $p$-orbital dominated antibonding bands.
In Fig. \ref{fig:BS}(a), the scalar-relativistic band structure [i.e., calculated without 
the spin-orbit coupling (SOC) included] is also displayed. Figure \ref{fig:BS}(a) shows that including
the SOC affects the band structure dramatically, as may be expected because bismuth has
the largest atomic number ($Z$) and thus has perhaps
the strongest SOC among the nonradioactive metals. For example, the gap between the top two valence bands
at both the X and F points increases from 0.12 eV to 1.29 eV when the SOC is switched-on. 

Interestingly, the $E_F$ falls at the pseudogap (zero energy).
Furthermore, the DOS increases steeply as energy ($E$) is lowered below $E_F$ but increases
much smoothly as $E$ is raised just above the $E_F$. This strong electron-hole asymmetry in the DOS 
spectrum around $E_F$ could be the main reason why bismuth has the large negative Seebeck coefficients of
$S_{xx} = -51$ $\mu$V/K and $S_{zz} = -103$ $\mu$V/K~\cite{Gallo1963}.

\begin{table*}
\caption{$R\overline{3}m$ symmetry-imposed shape of the spin Hall conductivity ($\sigma$) 
and spin Nernst conductivity ($\alpha$)
tensors~\cite{seeman2015,Gallego2019}. Here $Z = \sigma$ or $\alpha$.
Note that there are only four nonzero independent matrix elements, namely, $Z_{yx}^{z}$, $Z_{xz}^{y}$,
$Z_{zy}^{x}$ and $Z_{xx}^{y}$.
}
\begin{tabular}{c c c}
\hline\hline
\underline{$Z$}$^x$ & \underline{$Z$}$^y$ & \underline{$Z$}$^z$ \\
\hline
\\
$\left(\begin{array}{ccc} 0 & Z_{xx}^{y} & 0\\Z_{xx}^{y} & 0 & -Z_{xz}^{y}\\0 & Z_{zy}^{x}&0\end{array}\right)$ &
$\left(\begin{array}{ccc} Z_{xx}^{y} & 0 & Z_{xz}^{y}\\0 & -Z_{xx}^{y} & 0\\-Z_{zy}^{x} & 0&0\end{array}\right)$ &
$\left(\begin{array}{ccc} 0 & -Z_{yx}^{z} & 0\\Z_{yx}^{z} & 0 & 0\\0 & 0&0\end{array}\right)$ \\ \\
\hline\hline
\end{tabular}
\end{table*}

\begin{table*}
\caption{Calculated spin Hall conductivity (SHC) ($\sigma_{yx}^{z}$, $\sigma_{xz}^{y}$ and $\sigma_{zy}^{x}$),
and spin Nernst conductivity (SNC) ($\alpha_{yx}^{z}$, $\alpha_{xz}^{y}$ and $\alpha_{zy}^{x}$) 
at room temperature of Bi semimetal.
Previous results for some 5$d$ metals (Pt, Au and $\beta$-Ta)
are also listed for comparison.
To estimate spin Hall (Nernst) angle $\Theta_{sH} = 2\sigma^s/\sigma_{0}$
($\Theta_{sN}= 2\alpha^s/\alpha_{0} = \alpha^s/S_{0}/\sigma_{0}$), we also list experimental
electrical conductivity $\sigma_{0}$ and Seebeck coefficient $S_{0}$ as well as
the estimated $\Theta_{sH}$ and $\Theta_{sN}$. Note that the calculated SHC $\sigma_{xx}^{y}$ and
SNC $\alpha_{xx}^{y}$ are zero (within the numerical uncertainty) and thus are not listed here.
The energy derivatives of the $\sigma_{yx}^{z}$, $\sigma_{xz}^{y}$ and $\sigma_{zy}^{x}$ at the 
Fermi level are 200, -659, -276 ($\frac{\hbar}{e}\frac{1}{\Omega cm}$), respectively. 
}
\begin{ruledtabular}
\begin{tabular}{c c c c c c c c c c c}
System & $\sigma_{xx}$ ($\sigma_{zz}$) & $S_{xx}$ ($S_{zz}$) & $\sigma_{yx}^{z}$ & $\Theta_{sH}^z$ & $\sigma_{xz}^{y}$ ($\sigma_{zy}^{x}$) & $\Theta_{sH,xz}^y$ ($\Theta_{sH,zy}^x$) & $\alpha_{yx}^{z}$ & $\Theta_{sN}^z$ & $\alpha_{xz}^{y}$ ($\alpha_{zy}^{x}$) & $\Theta_{sN,xz}^y$ ($\Theta_{sN,zy}^x$) \\
     & ($\frac{10^4}{\Omega cm}$) & ($\frac{\mu V}{K}$) & ($\frac{\hbar}{e}\frac{1}{\Omega cm}$) & (\%) & ($\frac{\hbar}{e}\frac{1}{\Omega cm}$) & (\%) & ($\frac{\hbar}{e}\frac{A}{m K}$) & (\%) & ($\frac{\hbar}{e}\frac{A}{m K}$) & (\%) \\ \hline
Bi & 0.90 (0.75)$^{a}$ & -51 (-103)$^{a}$ & 1062, 474$^{b}$ & 24.0 & 903 (967) & 22 (23) & -0.139 & 0.61   & 0.349 (0.180) & -1.01 (-0.57) \\
   &            &           & $\sim$600$^{c}$& 0$\sim$24$^{d}$   &         &  0$\sim$24$^{d}$, 17$^{e}$  &        &    &                &  \\
   &            &           & 500$\sim$1000$^{f}$   &            &         &            &        &    &                &  \\
 Pt &20.8$^{g}$  & -3.7$^{h}$ & 2139$^i$ & 10$^{g}$ & --& -- & -1.09$^{j}$ & -20$^{h}$ & -- & -- \\
    &            &            & 2280$^k$ &          &   &    & -1.57$^{h}$ &           &    &    \\
 Au & 9.1$^{g}$  &  1.5$^{l}$ &  446$^m$ & 1.4$^{n}$ & --& -- & 0.13$^{o}$ & 1.8 & -- & -- \\
$\beta$-Ta &0.29$^{g}$  & -1.5$^{p}$ & -378$^m$, -389$^k$ &-1.0$^{n}$ & --& -- &  0.71$^{o}$ & -- & -- & -- \\
           &            &            & -400$^q$ &           &   &    &             &    &    &    \\
\end{tabular}
\end{ruledtabular}
$^{a}$Experiment at 300 K \cite{Gallo1963};
$^{b}$Tight-binding Hamiltonian calculation \cite{Sahin2015};
$^{c}$Experiment on Bi$_{1-x}$Sb$_x$ with $x \le 0.35$ \cite{Chi2020};
$^{d}$Experiments (see \cite{Yue2021} and references therein);
$^{e}$Experiment \cite{Fukumoto2022};
$^{f}$Experiment \cite{Chi2022};
$^{g}$Experiment \cite{Wang2014};
$^{h}$Experiment at 255 K \cite{Meyer2017};
$^{i}$\textit{Ab initio} calculation \cite{Guo2008};
$^{j}$\textit{Ab initio} calculation \cite{Guo2017};
$^{k}$\textit{Ab initio}-based Wannier interpolation \cite{Qiao2018};
$^{l}$Experiment \cite{Zolotavin2017};
$^{m}$\textit{Ab initio} calculation \cite{Guo2009,Qu2018};
$^{n}$Experiment at 300 K \cite{Qu2018};
$^{o}$This work; for computational details, see \cite{Qu2018};
$^{p}$Experiment at 300 K \cite{Fiflis2013};
$^{q}$Experiment \cite{Liu2012}.

\end{table*}

\subsection{Spin Hall effect}
As mentioned above, the SHC ($\sigma_{ij}^{s}$; $s,i,j=x,y,z$) for a material is a third-order tensor.
Nevertheless, for a highly symmetric crystal such as bismuth, most of the tensor elements are zero.
As mentioned before, bismuth has a rhombohedral $R\bar{3}$m structure with
point group $D_{3d}$.~\cite{Cucka1962}
In Table I, the shape of the SHC tensor for Bi metal,
uncovered by a symmetry analysis~\cite{seeman2015,Gallego2019}, is displayed.
Table I indicates that bismuth has only four nonzero independent elements,
namely, $\sigma_{yx}^{z}$, $\sigma_{xz}^{y}$, $\sigma_{zy}^{x}$ and $\sigma_{xx}^{y}$. 
Consequently, $\sigma_{yz}^{x} = -\sigma_{xz}^{y}$ and $\sigma_{zx}^{y} = -\sigma_{zy}^{x}$.
The results of our first principles calculations are consistent with this symmetry analysis
except that $\sigma_{xx}^{y}$ is found to be zero, i.e., we have only three independent
nonzero SHC elements ($\sigma_{yx}^{z}$, $\sigma_{xz}^{y}$ and $\sigma_{zy}^{x}$).
In Table II, all the three nonzero elements of the SHC of Bi are listed.
The SHCs of previously studied heavy metals Pt, Au and Ta
are also listed in Table II for comparison.
Table II indicates that all three nonzero SHC elements of Bi are large.
In particular, $\sigma_{yx}^{z}$ is 1062 ($\hbar$/e)(S/cm), being about half that 
of fcc Pt which possesses the largest intrinsic SHC [$\sim$2140 ($\hbar$/e)(S/cm)] 
among the elemental metals \cite{Guo2008,Guo2014}.
Furthermore, all three SHC elements are more than two times larger than that of Au and Ta (Table II).

Interestingly, Bi semimetal exhibits anisotropic SHE.
For an isotropic cubic metal (e.g., Pt), there is only one independent SHC element,
namely, $\sigma_{yx}^{z} = \sigma_{xz}^{y} = \sigma_{zy}^{x}$. In a hcp metal (e.g., Zr),
however, $\sigma_{yx}^{z} \ne  \sigma_{xz}^{y}$ although  $\sigma_{yx}^{z} = \sigma_{zy}^{x}$.~\cite{Freimuth2010}
Thus, one can define the SHC anisotropy as $\Delta_{zy}^{sH} = \sigma_{yx}^{z} - \sigma_{xz}^{y}$~\cite{Freimuth2010}.
For rhombohedral bismuth, $\sigma_{yx}^{z} \ne  \sigma_{xz}^{y} \ne  \sigma_{zy}^{x}$ (see Table I),
and consequently, the SHC anisotropy should be characterized by two parameters, namely,
$\Delta_{zy}^{sH}=\sigma_{yx}^{z}-\sigma_{xz}^{y}$ and $\Delta_{yx}^{sH}=\sigma_{xz}^{y}-\sigma_{zy}^{x}$.
Table II indicates that Bi semimetal shows significantly anisotropic spin Hall effect,
although its rhombohedral structure can be considered as a superposition of two very slightly distorted 
fcc lattices of Bi atoms~\cite{Cucka1962}. Specifically, Table II indicates that $\Delta_{zy}^{sH}=159$ ($\hbar$/e)(S/cm),
meaning that when the polarization of the spin current is switched from the $z$ direction to the $y$ direction,
the SHC would get reduced by about 15 \%. On the other hand, $\Delta_{yx}^{sH}=-64$ ($\hbar$/e)(S/cm)
suggest that the SHC would be increased by about 7 \% if the spin polarization is further rotated
from the $y$ axis to the $x$ axis.
As for the large anisotropic $g$ factor of holes in bismuth~\cite{Fuseya2015},
this significant crystalline anisotropy in the SHE of Bi semimetal could be attributed to its
peculiar band structure near the Fermi level and also its large SOC. 

We notice that for the application of SHE in spintronics such as spin-orbit torque
switching-based nanodevices~\cite{Hoffmann2013,Sinova2015}, the crucial quantity
is the so-called spin Hall angle $\Theta_{sH}$
which characterizes the charge-to-spin conversion efficiency.
The spin Hall angle is given by $\Theta_{sH} = (2e/\hbar)J^s/J^c = 2\sigma^s/\sigma^c$ where
$J^c$ and $\sigma^c$ are the longitudinal charge current density and conductivity, respectively
(see, e.g., Refs. \cite{Sinova2015} and \cite{Tung2013}).
Since bismuth has an electrical conductivity much smaller than that of Pt and Au (Table II),
we would expect that the $\Theta_{sH}$ values for all the nonzero SHC elements
should be much larger than Au and also larger than Pt.
Indeed, using the measured electrical conductivity for Bi,
we find that the $\Theta_{sH}$ values for $\sigma_{yx}^{z}$, $\sigma_{xz}^{z}$ and $\sigma_{zy}^{x}$
are 24 \%, 22 \% and 23 \%, respectively.
These values are significantly larger than that of Pt and Au (see Table II).

\begin{figure}[tbph] \centering
\includegraphics[width=8.0cm]{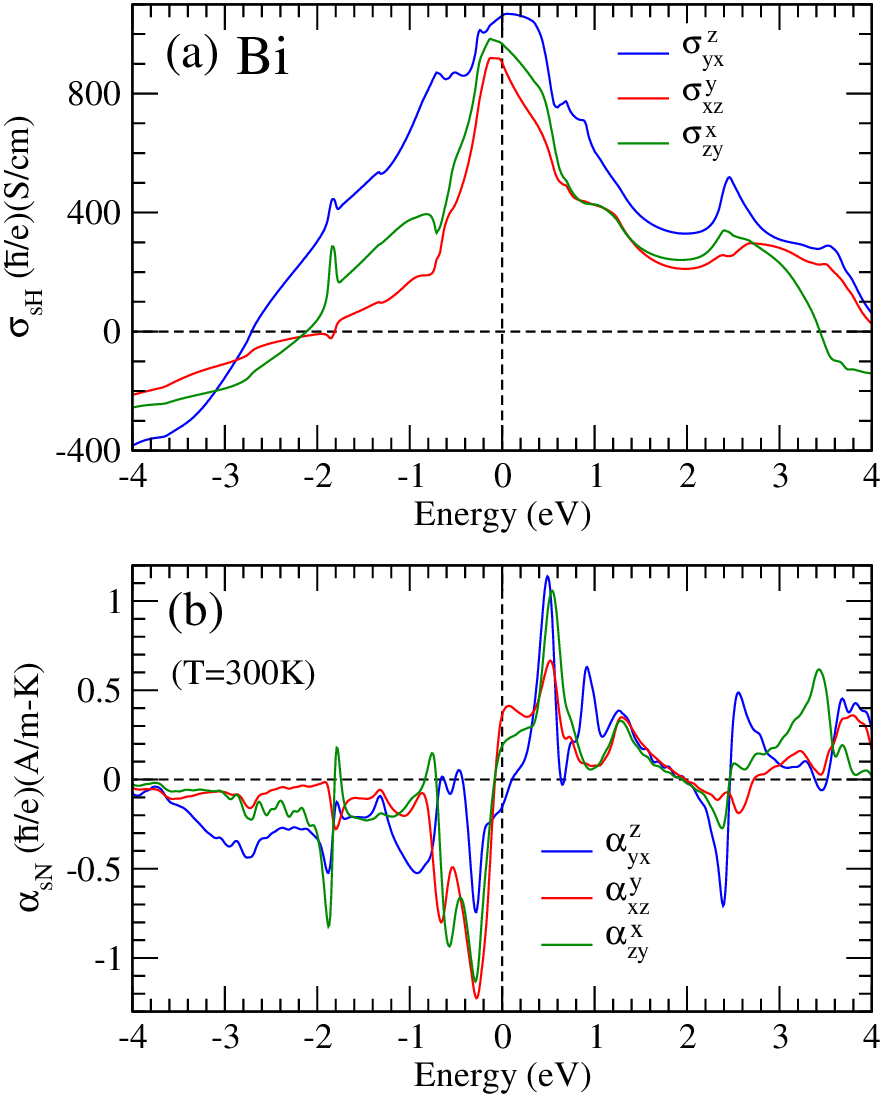}
\caption{(a) Spin Hall conductivity and (b) spin Nernst conductivity at $T=300$ K
as a function of the Fermi level of Bi semimetal.}
\label{fig:SHC}
\end{figure}

Since the SHC could be sensitive to the location of the Fermi energy ($E_F$)~\cite{Guo2005,Guo2008,Guo2014}, 
we also calculate the SHC as a function of $E_F$ within the so-called rigid band approximation, 
i.e., only the Fermi energy is varied while the band structure is kept fixed. 
The results may allow us to optimize the SHC of Bi by chemical doping or gating.
However, we should note that shifting the Fermi level by, e.g., chemical doping 
will inevitably change the band structure. If these changes in the band structure are large, 
they will render the rigid band approximation invalid. Nonetheless, if the amount of hole 
or electron doping is small, one may expect that the band structure is hardly affected
and hence the predictions are valid.
The obtained spectra of the SHC elements are displayed Fig. 2. 
Indeed, as one can see from Fig. 2, the SHC curves show a significant dependence on the $E_F$. 
In particular, all three elements have a prominent peak near the $E_F$ with the peak values being 
as large as about 1000 ($\hbar$/e)(S/cm) [see Fig. 2(a)].
This could be attributed to the fact that the Fermi level falls within the narrow band gaps
around the Z and L points in the BZ [Fig. 1(a)]. This is reminescent of platinum metal where
the SHC also exhibits a prominent peak at the $E_F$ which falls within the SOC-induced
band gaps~\cite{Guo2008}. It was pointed out in \cite{Guo2008} that when two degenerate
bands become slightly gapped by the SOC, a pair of large peaks of spin Berry curvature
with opposite signs would appear on the occupied and empty bands, respectively,
thus resulting in a gigantic contribution to the SHC [see Eq. (2)]. 
Indeed, Fig. 3(c) indicates that spin Berry curvatures $\Omega_{yx}^z$ 
of bands 3 and 4 peak in the vicinity of the Z and L points with large magnitudes but
opposite signs. Since band 3 is almost full while band 4 is nearly empty [Fig. 1(a)],
the total $\Omega_{yx}^z$ shows a gigantic positive peak near the Z point along the $\Gamma$-Z line
and also a smaller negative peak near the L point along the $\Gamma$-L line. 
Clearly, the large positive $\sigma_{yx}^z$ peak near the $E_F$ results from the
gigantic $\Omega_{yx}^z$ near the Z point [Fig. 1(c)].

As the energy is either raised above the $E_F$ or lowered below the $E_F$, 
the magnitudes of all three elements decrease monotonically [Fig. 2(a)]. 
Interestingly, the anisotropy in the SHE gets considerably enhanced.
For example, when the $E_F$ is shifted upwards by 0.20 eV,
the magnitudes of the $\Delta_{zy}^{sH}$ and $\Delta_{yx}^{sH}$
would almost be doubled, changing to 280 ($\hbar$/e)(S/cm) and -120 ($\hbar$/e)(S/cm), respectively.
This can be easily achieved by electron doping of 0.009 e/unit cell (or $n_e = 1.3\times10^{20}$ cm$^{-3}$) 
via substitutional alloying Bi$_{1-x}$Te$_{x}$~\cite{Chi2022}.
Similarly, if the $E_F$ is lowered by 0.35 eV,
the magnitudes of the $\Delta_{zy}^{sH}$ and $\Delta_{yx}^{sH}$
would also be nearly doubled, changing to 289 ($\hbar$/e)(S/cm) and -133 ($\hbar$/e)(S/cm), respectively.
This can be accomblished by hole doping of 0.190 e/unit cell (or $n_h = 2.7\times10^{21}$ cm$^{-3}$)
via substitutional alloying Bi$_{1-x}$Sn$_{x}$~\cite{Chi2022}.
At around -2.0 eV, $\sigma_{xz}^y$ and $\sigma_{zy}^x$ change sign while $\sigma_{yx}^z$
changes sign at about -2.7 eV.
Figure 2(a) shows that in the energy range from -2.0 to -0.5 eV,
there is a prominent difference between $\sigma_{xz}^y$ and $\sigma_{yx}^z$,
thus indicating a strongly anisotropic SHE.
For example, $\Delta_{zy}^{sH}$ is as large as 637 ($\hbar$/e)(S/cm) at -0.72 eV [Fig. 2(a)]. 

As mentioned before, one group reported the realistic tight-binding Hamiltonian calculation of
the intrinsic SHC for Bi. Nevertheless, only one independent element ($\sigma_{yx}^{z}$)
was calculated.~\cite{Sahin2015} Overall, the shape of the $\sigma_{yx}^{z}(E)$ spectrum reported 
in ~\cite{Sahin2015} agrees well with that displeyed in Fig. 2(a).  
However, Table II indicates that the $\sigma_{yx}^{z}$ value [472 ($\hbar$/e)(S/cm)]
reported in ~\cite{Sahin2015} is more than two times smaller than that [1062 ($\hbar$/e)(S/cm)]
of the present calculation. We are unable to understand the origin of this
big discrepancy between the two calculations since we do not know the details of the previous calculation~\cite{Sahin2015}.
Nevertheless, we do notice that our calculated SHC values for fcc Pt and $\beta$-Ta
(Table II) agree very well with the previous independent calculations
using the different methods~\cite{Guo2008,Qiao2018,Ryoo2019} (see Table II).

As noted above, the magnitude of the reported $\Theta_{sH}$ values varies from zero (negligibly small)
to 24 \%, depending on the spin current detection method and also the magnetic material used to inject
spin current in the experiments~\cite{Hou2012,Emoto2016,Zhang2015,Yue2018,Yue2021,Sangiao2021,Fukumoto2022}.
Such a wide range of the measured $\Theta_{sH}$ values could also suggest that there may be 
significant contributions from the extrinsic mechanisms
such as the side-jump and skew scattering~\cite{Hoffmann2013,Sinova2015}, which would depend on
the quality of the samples used in the experiments.
Also, we notice that the Bi films used in the experiments were deposited on different substrates
and consequently, the substrate-dependent charge transfer may occur.
Because both the magnitude and anisotropy of the SHC are sensitive to carrier doping [see Fig. 2(a)],
this could be another reason why the reported $\Theta_{sH}$ values differ significantly from each other.
Since the spin Hall angle also depends on the electric conductivity, it would be better
to compare the experimentally determined and theoretically calculated SHC.
In this context, we notice that the measured SHC for Bi$_{1-x}$Sb$_x$ with $x \le 0.35$
is ~600 ($\hbar$/e)(S/cm)~\cite{Chi2020}. Very recently, the SHC of 500$\sim$1000 ($\hbar$/e)(S/cm)
in pure bismuth was also reported~\cite{Chi2022}. These experimental SHC values 
are quite close to our theoretical SHC values (Table II).  
Furthermore, the SHC of the doped bismuth as a function of carrier concentration
was reported~\cite{Chi2022}, which exhibits a pronounced peak
at pure bismuth, agreeing well with the SHC versus $E_F$ curves displayed in Fig. 2(a).
We hope that our interesting finding of anisotropic and tunable spin Hall conductivity and angle
would spur further experiments on the SHE in bismuth semimetal using
single crystal specimens.
 
\subsection{Spin Nernst effect}
The SNC ($\alpha_{ij}^{s}$) for a solid is also a third-order tensor.
Since the transverse spin currents generated by a longitudinal electric field 
and a longitudinal temperature gradient have the same transformation properties under the symmetry operations,
the shape of the SNC is identical to that of the SHC \cite{seeman2015} (see Table I).  
Therefore, the SNC of bismuth semimetal also has only three nonzero independent elements, namely, 
$\alpha_{yx}^{z}$, $\alpha_{xz}^{y}$, and $\alpha_{zy}^{x}$.
Table II shows that the calculated values of these nonzero elements of the SNC tensor
$T = 300$ K are rather significant. 
In fact, they are in the same order of magnitude as that of gold metal (Table II).
Nevertheless, they are a few times smaller than that of platinum metal.

Interestingly, Table II indicates that the SNC of Bi is more anisotropic than the SHC.
First of all, in contrast to $\sigma_{yx}^{z}$ and $\sigma_{zy}^{x}$,
$\alpha_{yx}^{z}$ and $\alpha_{xz}^{y}$ have opposite signs. 
The magnitude of $\alpha_{xz}^{y}$ is nearly three times larger than that of $\alpha_{yx}^{z}$.
This gives rise to the large anisotropic parameter 
$\Delta_{zy}^{sN} = \alpha_{yx}^{z} - \alpha_{xz}^{y} = 0.49$ ($\hbar$/e)(A/m-K),
indicating that when the spin polarization is switched from the $z$ axis to
the $y$ axis, the SNC would not only change sign but also get enhanced by a factor of $\sim$3.
On the other hand, we notice that $\alpha_{zy}^{x} = 0.18$ ($\hbar$/e)(A/m-K),
meaning that the SNC would be reduced by half if the spin polarization is further rotated from
the $y$ axis to the $x$ axis. 

As for the SHC, we also calculate the three elements of the SNC as a function of $E_F$.
In Fig. 2(c), all the three SNC elements ($\alpha_{yx}^{z}$, $\alpha_{xz}^{y}$ and $\alpha_{zy}^{x}$)
at $T = 300$ K are displayed as a function of $E_F$. 
Furthermore, we note that in the low temperature limit, Eq. (\ref{eq:3})
can be written as the Mott relation,
\begin{equation}
\begin{aligned}
\alpha_{ij}^{s}(E_F)=-\frac{\pi^2}{3}\frac{k_B^2T}{e}\sigma_{ij}^{s}(E_F)',
\end{aligned}
\end{equation}
which indicates that the SNC is proportional to the energy derivative of the SHC
at the $E_F$. In other words, a peak in the SNC would occur when the SHC
has a steep slope. Interestingly, the Mott relation shows the reason that $\alpha_{yx}^{z}$
is negative while $\alpha_{xz}^{y}$ and $\alpha_{zy}^{x}$ are positive [see Table II
and Fig. 2(b)], is because the slope of $\sigma_{yx}^{z}$ is positive while 
the slopes of $\sigma_{xz}^{y}$ and $\sigma_{zy}^{x}$ are negative [see Fig. 2(a)].

Figure 2(c) clearly shows that the SNC has a stronger $E_F$ dependence than the SHC.
Remarkably, both $\alpha_{xz}^{y}$ and $\alpha_{zy}^{x}$ decreases sharply 
as the energy is lowered from the $E_F$ and change sign at -0.08 eV and then
reaches its minimum of -1.22 ($\hbar$/e)(A/m-K) at -0.28 eV 
when the energy is further lowered [see Fig. 2(c)].
Note that hole doping of merely 0.001 and 0.108 e per unit cell, respectively, 
would realize the sign change and also reach the maximal magnitude. 
The Mott relation clearly indicates that this is caused by the rapid change
of the slope of the corresponding SHC elements in this energy region.
On the other hand, when the energy increases, the $\alpha_{xy}^{z}$
increases steadily, changes sign at $\sim$0.11 eV and then reaches 
its maximum of 1.14 ($\hbar$/e)(A/m-K) at 0.49 eV, which can be obtained by
electron-doping of 0.122 e per unit cell. 
Note again that the peak is located at the point where $\sigma_{xy}^{z}$ drops sharply [Fig. 2(a)],
being consistent with the MOtt relation.

\begin{figure}[tbph] \centering
\includegraphics[width=8.0cm]{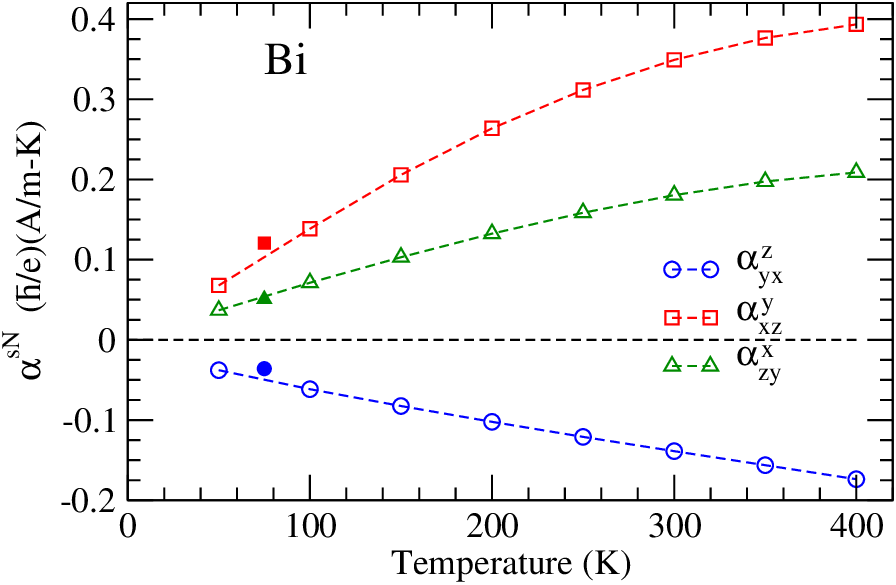}
\caption{Spin Nernst conductivity (SNC) ($\alpha_{yx}^{z}$, $\alpha_{xz}^{y}$ and $\alpha_{zy}^{x}$) 
of Bi semimetal as a function of temperature. The solid symbols denote the 
SNC values estimated using the energy derivatives of the SHC at the $E_F$ (see Table II) 
and the Mott relation [Eq. (4)]. }
\label{fig:T-dep}
\end{figure}

We also calculate the three independent elements of the SNC as a function of $T$,
as displayed in Fig. \ref{fig:T-dep}. 
Figure \ref{fig:T-dep} shows that $\alpha_{xz}^{y}$ and $\alpha_{zy}^{x}$ are positive 
while $\alpha_{yx}^{z}$ is negative in the entire considered temperature range.
The magnitude of all the three elements increases monotonically with $T$.
In the low temperature region (e.g., $T \le 200$ K), all the elements are roughly
proportional to $T$, as indicated by the Mott relation [Eq. (4)]. 
Using the calculated energy derivatives of the SHC elements (Table II) and
Eq. (4), we can estimate the SNC elements at a low temperature.
Indeed, Fig. \ref{fig:T-dep} indicates that such estimated values of the SNC elements at 75 K
agree quite well with that of the full calculation using Eq. (3),
verifying that the Mott relation holds quantitatively at low temperatures.

Finally, from the viewpoint of the application of SNE in spin caloritronics, the key quantity is
the spin Nernst angle $\Theta_{sN}$ which measures the heat-to-spin conversion efficiency and
is given by $\Theta_{sN} = (2e/\hbar)J^s/J^h = 2\alpha^s/\alpha^L$ where
$J^h$ and $\alpha^L$ are the longitudinal heat current density and Nernst coefficient, respectively~\cite{Meyer2017}.
Here $\alpha^L = S_{aa}\sigma_{aa}$ where $S_{aa}$ is the Seebeck coefficient.
Using the measured $\sigma_{xx}$ ($\sigma_{zz}$) and $S_{xx}$ ($S_{zz}$) as well as
the calculated $\alpha^s$ of bismuth semimetal (Table II), 
we estimate the $\Theta_{sN}$ values and list the results in Table II. 
The obtained $\Theta_{sN}$ values are around 1 \%, which are significant.
Nevertheless, they are more than ten times smaller than that (-20 \%)
of Pt metal~\cite{Meyer2017}. This is because not only the Seebeck coefficients of Bi semimetal
are much larger than that of Pt metal but also the calculated SNC values of Bi are several times smaller than
that of Pt (see Table II).

\section{CONCLUSIONS}
Summarising, we have performed a thorough theoretical study of the SHE and SNE in bismuth
based first-principles relativistic band structure calculations 
and the Berry phase formalism. Our symmetry analysis indicates that unlike cubic metals such as
Pt and Au, the SHC and SNC tensors have three indepedent nonzero elements,
viz., $Z_{yx}^z$, $Z_{xz}^y$ and $Z_{zy}^x$.
We have calculated all the three elements as a function of the Fermi energy ($E_F$),
instead of one element in the previous calculation~\cite{Sahin2015}.
The results of these calculations show that all SHC tensor elements are large, 
being $\sim$1000 ($\hbar$/e)(S/cm). They are comparable to that of platinum which posseses the
largest intrinsic SHC among the metals. 
Furthermore, because bismuth semimetal has low electrical conductivity, 
the corresponding spin Hall angles are gigantic, being $\sim$20 \% (see Table II)
and much larger than that of platinum (Table II).
All the calculated SNC tensor elements are also pronounced, being
comparable to that [$\sim$0.13 ($\hbar$/e)(A/m-K)] of gold, 
although they are several times smaller than that of platinum.
Both the magnitude and sign of all the calculated conductivity elements 
vary strongly with the Fermi level (Fig. 2),
indicating that the SHE and SNE in bismuth could be opitmized 
by varying the Fermi level via, e.g., chemical doping. 
All these suggest that bismuth is a promising material for spintronics and spin caloritronics.
Interestingly, in contrast to Pt and Au where $Z_{yx}^z = Z_{xz}^y = Z_{zy}^x$,
the SHE and SNE in bismuth are anisotropic,
i.e., $Z_{yx}^z$, $Z_{xz}^y$ and $Z_{zy}^x$ differ significantly.
In particular,  $\alpha_{yx}^z$ and  $\alpha_{xz}^y$ even differ in sign.
Consequently, the Hall voltages due to the inverse SHE and inverse SNE from the different
conductivity elements could cancel each other and thus result
in a small spin Hall angle if polycrystalline samples are used, which
may explain why the measured spin Hall angles ranging from nearly 0 to 25 \% have been reported.
We thus believe that this work would stimulate further spin current experiments on bismuth
using highly oriented single crystal specimens.


\section*{ACKNOWLEDGMENTS}
The author thanks Prof. Chia Ling Chien for many enlightning discussions and advices on
magnetism, superconductivity and spin-related transports in general as well as 
on the fascinating properties of bismuth semimetal in particular over the years.
The author acknowledges the support from the Ministry of Science and Technology 
and the National Center for Theoretical Sciences, Taiwan.



\end{document}